\def\virg#1{``#1"}

\def\eqi{\begin{equation}}
\def\eqf{\end{equation}}
\def\eqia{\begin{eqnarray}}
\def\eqfa{\end{eqnarray}}

\def\ton#1{\left(#1\right)}

\documentclass[onecolumn]{aastex}
\usepackage{morefloats}
\usepackage[title]{appendix}
\usepackage{hyperref}
\usepackage{booktabs}
\usepackage[table,xcdraw]{xcolor}
\usepackage{multirow}
\usepackage{rotating,tabularx}
\usepackage{float}
\usepackage{enumerate}
\usepackage{rotating}
\usepackage[polutonikogreek,english]{babel}
\usepackage{amsmath,starfont,textgreek,w-greek,wasysym}
\usepackage{amsthm}
\usepackage{amscd,lineno}
\usepackage{amssymb,dsfont}
\usepackage{graphicx,epsfig}
\usepackage{txfonts}
\usepackage{xr-hyper}

\RequirePackage{color}

\newcommand{\emaila}{lorenzo.iorio@libero.it}

\allowdisplaybreaks[1]

\begin{document}

\title{A comment on \virg{Can observations inside the Solar System reveal the gravitational properties of the quantum vacuum?} by D.S. Hajdukovic}

\shortauthors{L. Iorio}

\author{Lorenzo Iorio\altaffilmark{1} }
\affil{Ministero dell'Istruzione, dell'Universit\`{a} e della Ricerca
(M.I.U.R.)-Istruzione
\\ Permanent address for correspondence: Viale Unit\`{a} di Italia 68, 70125, Bari (BA),
Italy}

\email{\emaila}

\begin{abstract}
The modified gravitational theory by Hajdukovic, based  on the idea that quantum vacuum  contains virtual gravitational dipoles, predicts, among other things, anomalous secular precessions of the planets of the Solar System as large as $\simeq 700-6,000$ milliarceconds per century. We demonstrate that they are ruled out by several orders of magnitude by the existing bounds on any anomalous orbital secular rates obtained with the EPM and INPOP ephemerides.
\end{abstract}

keywords{
ephemerides - celestial mechanics  - gravitation - planets and satellites: dynamical evolution and stability
}
\section{Introduction}
Some years ago, \citet{utrmon} put forth the hypothesis that quantum vacuum would contain virtual gravitational dipoles. He argued that this
hypothesis, which would have the potential to simultaneously solve the Dark Matter and Dark Energy problems, might be tested within the Solar System. The key point in his proposal  consists of the fact that the quantum vacuum (\virg{enriched} with the gravitational dipoles) would induce a retrograde precession of the perihelion because of an additional constant radial acceleration of gravitational origin.
In his Table\,1, \citet{utrmon}  calculated the amount of such a putative anomalous precession for the planets of the Solar System finding values ranging from $-690$ milliarcseconds per century $\ton{\mathrm{mas\,cty}^{-1}}$ for Mercury to $-5,980\,\mathrm{mas\,cty}^{-1}$ for Neptune. \citet{utrmon} did not compare his theoretical predictions with the very tight bounds, already existing at the time of his writing, on any anomalous perihelion precessions for the inner planets of the Solar System and Saturn released by teams led by the astronomers E.V. Pitjeva and A. Fienga. Instead, he decided to limit himself just to Uranus and Neptune by citing, e.g., a paper of the present author, published in the context of the Pioneer Anomaly and relying upon \citet{2005SoSyR..39..176P} who used the EPM2004 ephemerides, by stating that \virg{the current ephemerides of planets do not preclude the illustrative values} of his Table\,1. In fact, the bounds on the perihelion precessions of the outer planets of the Solar System \citep{Pit010} were and, to our knowledge, are still too weak  to rule out with confidence even effects as large as those listed by \citet{utrmon} in his Table\,1. As a result, \citet{utrmon} suggested to look at the pericenter precession of the orbital motion of  the natural satellite Dysnomia around its primary which is the dwarf planet Eris.
\section{A comparison with the EPM and INPOP ephemerides}
Actually, the anomalous effects predicted by \citet{utrmon} for the telluric planets should have been deemed as completely incompatible even with the planetary data processed with the EPM2004 ephemerides and available since 2005. Indeed, Table\,3 of \citet{2005AstL...31..340P} states that the uncertainties in the perihelion precessions of the inner planets were as little as $5\,\mathrm{mas\,cty}^{-1}$ (Mercury), $0.4\,\mathrm{mas\,cty}^{-1}$ (Earth), and $0.5\,\mathrm{mas\,cty}^{-1}$ (Mars); Table\,1 of \citet{utrmon} predicts anomalous precessions as large as about $\simeq -1,000\,\mathrm{mas\,cty}^{-1}$ for the Earth and Mars. Later, the situation became even worse. Suffice it to say that the INPOP10a ephemerides allowed \citet{2011CeMDA.111..363F} to obtain bounds on the inner planets of the Solar System as tiny as $0.6\,\mathrm{mas\,cty}^{-1}$ (Mercury), $0.9\,\mathrm{mas\,cty}^{-1}$ (Earth), and $0.15\,\mathrm{mas\,cty}^{-1}$ (Mars), while \citet{2013AstL...39..141P} and \citet{2013MNRAS.432.3431P}  obtained  $3\,\mathrm{mas\,cty}^{-1}$ (Mercury), $0.19\,\mathrm{mas\,cty}^{-1}$ (Earth), and $0.037\,\mathrm{mas\,cty}^{-1}$ (Mars) with the EPM2011 ephemerides. Moreover, while Table\,1 of \citet{utrmon} predicts a perihelion precession of $-3,360\,\mathrm{mas\,cty}^{-1}$ for Saturn, the INPOP10a \citep{2011CeMDA.111..363F} and EPM2011 \citep{2013AstL...39..141P,2013MNRAS.432.3431P} ephemerides yielded uncertainties of  $0.65\,\mathrm{mas\,cty}^{-1}$ and $0.47\,\mathrm{mas\,cty}^{-1}$, respectively.
\section{Conclusions}
Thus, we conclude that the exotic secular rates of change predicted by \citet{utrmon} are neatly ruled out by the planetary observations, and not even the most recent ones.

\bibliography{MS_binary_pulsar_bib,Gclockbib,semimabib,PXbib}{}

\begin{thebibliography}{7}
\expandafter\ifx\csname natexlab\endcsname\relax\def\natexlab#1{#1}\fi

\bibitem[{{Fienga} {et~al}\mbox{.}(2011){Fienga}, {Laskar}, {Kuchynka},
  {Manche}, {Desvignes}, {Gastineau}, {Cognard}, \&
  {Theureau}}]{2011CeMDA.111..363F}
{Fienga} A., {Laskar} J., {Kuchynka} P., {Manche} H., {Desvignes} G.,
  {Gastineau} M., {Cognard} I., {Theureau} G., 2011, Celest. Mech. Dyn. Astr.,
  111, 363

\bibitem[{{Hajdukovic}(2013)}]{utrmon}
{Hajdukovic} D., 2013, Astrophys. Space Sci., 343, 505

\bibitem[{{Pitjev} \& {Pitjeva}(2013)}]{2013AstL...39..141P}
{Pitjev} N.~P., {Pitjeva} E.~V., 2013, Astronomy Letters, 39, 141

\bibitem[{{Pitjeva}(2005{\natexlab{a}})}]{2005SoSyR..39..176P}
{Pitjeva} E.~V., 2005{\natexlab{a}}, Solar Syst. Res., 39, 176

\bibitem[{{Pitjeva}(2005{\natexlab{b}})}]{2005AstL...31..340P}
{Pitjeva} E.~V., 2005{\natexlab{b}}, Astron. Lett., 31, 340

\bibitem[{{Pitjeva}(2010)}]{Pit010}
{Pitjeva} E.~V., 2010, in Proceedings of the International Astronomical Union,
  Vol. 261, Relativity in Fundamental Astronomy, {Klioner} S.~A., {Seidelmann}
  P.~K., {Soffel} M.~H., eds., pp. 170--178

\bibitem[{{Pitjeva} \& {Pitjev}(2013)}]{2013MNRAS.432.3431P}
{Pitjeva} E.~V., {Pitjev} N.~P., 2013, MNRAS, 432, 3431

\end{thebibliography}

\end{document}